\documentclass[12pt]{article}

\usepackage{epsfig}

\begin{document}

\newcommand{\beq}{\begin{equation}}
\newcommand{\eeq}{\end{equation}}
\newcommand{\beqa}{\begin{eqnarray}}
\newcommand{\eeqa}{\end{eqnarray}}

\def\ov{\overline}
\def\onlyif{\rightarrow}

\def\openone{\leavevmode\hbox{\small1\kern-3.8pt\normalsize1}}

\def\a{\alpha}
\def\b{\beta}
\def\g{\gamma}
\def\r{\rho}
\def\minus{\,-\,}
\def\eks{\bf x}
\def\kay{\bf k}

\def\ket#1{|\,#1\,\rangle}
\def\bra#1{\langle\, #1\,|}
\def\braket#1#2{\langle\, #1\,|\,#2\,\rangle}
\def\proj#1#2{\ket{#1}\bra{#2}}
\def\expect#1{\langle\, #1\, \rangle}
\def\trialexpect#1{\expect#1_{\rm trial}}
\def\ensemblexpect#1{\expect#1_{\rm ensemble}}
\def\kpsi{\ket{\psi}}
\def\kphi{\ket{\phi}}
\def\bpsi{\bra{\psi}}
\def\bphi{\bra{\phi}}

\def\ditto{\rule[0.5ex]{2cm}{.4pt}\enspace}
\def\th{\thinspace}
\def\ni{\noindent}
\def\thirty{\hbox to \hsize{\hfill\rule[5pt]{2.5cm}{0.5pt}\hfill}}

\def\set#1{\{ #1\}}
\def\setbuilder#1#2{\{ #1:\; #2\}}
\def\Prob#1{{\rm Prob}(#1)}
\def\pair#1#2{\langle #1,#2\rangle}
\def\Id{\bf 1}

\def\dee#1#2{\frac{\partial #1}{\partial #2}}
\def\deetwo#1#2{\frac{\partial\,^2 #1}{\partial #2^2}}
\def\deethree#1#2{\frac{\partial\,^3 #1}{\partial #2^3}}

\newcommand{\xx}{{\scriptstyle -}\hspace{-.5pt}x}
\newcommand{\yy}{{\scriptstyle -}\hspace{-.5pt}y}
\newcommand{\zz}{{\scriptstyle -}\hspace{-.5pt}z}
\newcommand{\kk}{{\scriptstyle -}\hspace{-.5pt}k}
\newcommand{\sx}{{\scriptscriptstyle -}\hspace{-.5pt}x}
\newcommand{\sy}{{\scriptscriptstyle -}\hspace{-.5pt}y}
\newcommand{\sz}{{\scriptscriptstyle -}\hspace{-.5pt}z}
\newcommand{\sk}{{\scriptscriptstyle -}\hspace{-.5pt}k}

\def\openone{\leavevmode\hbox{\small1\kern-3.8pt\normalsize1}}

\title{Rivisiting Token/Bucket Algorithms in New Applications}
\author{Andrea Pasquinucci
\\
\small
{\it {\rm UCCI.IT}, via Olmo 26, I-23888 Rovagnate (LC), Italy 
}}
\date{June 02, 2009}
\maketitle

\abstract{We consider a somehow peculiar Token/Bucket problem
which at first sight looks confusing and difficult to solve.
The winning approach to solve the problem consists in 
going back to the simple and traditional methods to solve
computer science problems like the one taught to us by Knuth.
Somehow the main trick is to be able to specify clearly 
what needs to be achieved, and then the solution, even if
complex, appears almost by itself.} 
\vspace{1 cm} 
\normalsize

\section{Introduction}

In designing computer programs it often happens to have to implement 
subtle logics which are confusing and can look difficult to realize. But 
most of the times the problem can be solved even in simple ways. The 
main point of this article is that often it is better to go back to 
basic, try hard to understand what it is that we need to implement and 
design a little algorithm following even old and traditional thinking
\cite{Knuth}.

We will describe here a case of a somehow peculiar token/bucket
algorithm, which at the end is just another variation on a infinite 
theme. The key to the solution of the problem, which can be made 
mathematically sound, is just a little reasoning by examples trying to 
isolate the important aspects of the problem.

The main issue is to formalize the problem in such clear terms that we 
can solve it. We'll see that once the problem has been described in a 
simple and precise way, the solution will appear almost automatically to 
us.

The problem here described arised in designing an application in a 
distributed environment. Today more and more often massive computing is  
approached by using many (relatively) inexpensive machines and 
by running programs in parallel on them. We are getting used to hear about
{\sl Cloud Computing}, {\sl share-nothing clusters} and {\sl distributed 
computing architecture based on parallel processing} but we need to
prepare our data and write our program in such a way as to make the 
best use of these new architectures. 

A typical problem incurred in these architectures is to split 
the input data so to be processed in parallel. The processing of
the data usually happens in steps, and splitting and merging
of data can occur various times during the processing. 

To optimize the distribution of the computational load and the
minimization of the transfer of data between nodes (data transfer
between physically distinct nodes is always a very slow process
with respect to in memory or even on disc processing), we need
to be very careful on how to split the data initially between the
members of the cluster. 

A possibile formalization of this processing architecture, the one 
useful and used here, is to represent the processing of data as
happening inside buckets and the data itself as tokens. 
So a cluster of machines or a group of parallel processes, 
is represented by a set of buckets in which we distribute our tokens,
that is the data to be processed. The various steps of the processing
can be formalized by different sets of buckets between which we move
the tokens. Our aim is to optimize the initial splitting of the 
data/tokens in the buckets.

\section{A first description of the problem}

The problem at hand consists in a three steps processing where
the first two steps happen in the same set of buckets, that is 
cluster of machines or group of parallel processes, but with
an important specification, and the third step happens 
in a second set of buckets. The tokens are nothing else than
input data which can be splitted to be processed in parallel.

Actually to optimize the splitting of the data/tokens in the
first set of buckets, the tokens are first splitted in a subset
of the first set of buckets and only in the second step they 
are distributed in all buckets of the first set. This peculiar
process depends on the details of the processing of the data,
and it is not relevant to the algorithmic problem we want to describe.

The setup is then the following. We have initially $T$ tokens and we 
need to design an algorithm to put the tokens in $B$ buckets first
and in another set of $B'$ buckets later.
Each token has also a {\sl label} which is related to the number of the token 
itself and which we will use to select the bucket where to put the 
token. There are a few constraints:

\begin{itemize}
\item we should initially put the tokens only in $C$ consecutive buckets 
($C \leq B$) out of the $B$ buckets of the first set, 
and the first bucket of the $C$ set can 
be any bucket in $B$ (so we count the $B$ buckets $mod B$ as if they make
up a ring)

\item the tokens should be distributed homogeneously in the buckets both 
with respect to the number of tokens in the $C$ buckets and to the 
distribution of the labels in the $C$ buckets

\item in the second step, the tokens will be redistributed in the $B-C$ buckets 
which have been left out in the first step, with the constraint that each 
token can be moved only once and that tokens cannot be redistributed 
between the $C$ buckets already filled in, still the final distribution 
of tokens must remain homogeneous between all buckets both in number and 
label

\item finally the tokens will be moved to the second set of $B'$ buckets 
with $B'>B$ and they must be distributed homogeneously without changing 
the label of each token. 
\end{itemize}

\section{Discussing examples}

Even if this formulation of the problem gives us an idea of what we need 
to do, it is not precise enough to permit to find an algorithm which 
solves it. To solve the problem we need a much more precise description 
of the problem, and for that we need to go back to what we want to 
realize and describe it in more details.

The best way is to try first with a couple of examples.

The first case is when $C=1$, this is an exceptional case where all 
tokens go in the same bucket and it does not teach us much. Anyway we 
still have to keep in mind that our final algorithm must work also when 
$C=1$.

The first interesting case is instead when $C=B$. We can start to put 
the first token in the first bucket and continue like that ($mod B$). 
This is a 
round-robin distribution. In this case we do not need to redistribute 
the tokens in the first set of buckets since they will be all already 
full. 

Notice that the number of tokens in each bucket differs at most by 1 
from the number of tokens in each other bucket. We can say that this is 
our homogeneity property with respect to the number of tokens in the 
buckets.

The first attempt to define the label is to simply set it to the number 
of the bucket where we put the token. This anyway will not work when we 
will put the tokens in the second set of buckets since we have $B'$ 
buckets in this case. Instead we can set $l(t)=t$ where $l$ is the label 
and $t$ is the number of the token, and choose the bucket where to put 
a token with $l(t)\ mod B$ for the first set of buckets, and 
$l(t)\ mod B'$ for the second set of buckets.

But we could start to put the tokens not in the first bucket, but in any 
bucket. Let denote by $f$ the first bucket in the first set of buckets 
where we put the first token. Then we need to modify the $l$ map to make 
it work when $f \neq 0$, but the solution is quite simple: $l(t)=t+f$. 
The maps to select the bucket where to put each token are still 
$l(t)\ mod B$ for the first set of buckets, and $l(t)\ mod B'$ 
for the second set of buckets.

Now what to do when $1 < C < B$ ? we can start by assigning $l(t)=t+f$ 
and put the first $C$ tokens as before, this works fine. The problem 
comes with the next $B-C$ tokens, where do we put them? The map 
$l(t)\ mod B$ does not work because it selects a bucket not in the set of 
the $C$ buckets to fill in at first. 
On the other side, the definition of $l$ and the two 
maps to select the bucket where to put a token, will work fine when we 
will have to move the buckets to the $B-C$ empty buckets of the first 
set, and later on to the $B'$ buckets of the second set. So we keep $l$ 
and the two maps as they are. 

What we can do in this case instead is to have two independent 
round-robin cycles, one for the tokens where $l(t)\ mod B$ is one of the 
buckets we should fill in, and one for the other tokens. The two 
round-robin cycles run independently and fill in the same buckets. If 
the two round-robin cycles run in the same direction, at the end we can 
have buckets with number of tokens which differ by 2 from the number of 
tokens in other buckets (if this it is not obvious, we will see it in 
details later on). But there is a simple way to avoid this, which is to 
run the two round-robin cycles in opposite directions, one clockwise 
and the other counter-clockwise (assuming that the buckets make a circle
$mod C$, that is they form a ring).

\section{A formal description}

Now the description of the algorithm is getting a little complicated, so 
we need to formalize it to make it explicit and to be able to prove the 
correctness of its solution.

We can then try with the following formal description of the problem.

Given an ordered set ${\cal B}=\{b_0,b_1 \dots b_{B-1}\}$ of $B$ 
buckets, 
an ordered set ${\cal B}'=\{b'_0,b'_1 \dots b'_{B'-1}\}$ of $B'>B$
buckets, 
an ordered subset ${\cal C}=\{b_f,b_{f+1\ mod B} \dots b_{f+C-1\ mod B}\}$ 
of ${\cal B}$ with $C$ buckets, 
and an ordered set ${\cal T}=\{t_0,t_1 \dots t_{T-1}\}$ of $T$ of tokens, 
let ${\cal I}_{\cal B}$, ${\cal I}_{{\cal B}'}$, ${\cal I}_{\cal C}$ and 
${\cal I}_{\cal T}$ 
be the ordered set of integers given by the indices of the elements 
of ${\cal B}$, ${\cal B}'$, ${\cal C}$ and ${\cal T}$ respectively.

Assign to each token $t_i$ a label $l$ by the map $L: {\cal T} \rightarrow 
{\cal L}$ where ${\cal L}$ is isomorphic to ${\cal T}$ and $l: {\cal 
I}_{\cal T} \rightarrow {\cal I}_{\cal L} $. In practice we choose ${\cal 
I}_{\cal L} $ to be the same set of integers as ${\cal I}_{\cal T} $ up to a 
constant shift $f$ and a little empty interval towards the high end of the set, 
as described below.

Let $M$ be a map $M : {\cal T} \rightarrow {\cal C}$, $m$ the associated map 
$m : {\cal I}_{\cal L} \rightarrow {\cal I}_{\cal C}$, $M' : {\cal T} \rightarrow 
{\cal B}'$, $m' : {\cal I}_{\cal L} \rightarrow {\cal I}_{{\cal B}'}$, $M'' : 
{\cal T} \rightarrow {\cal B}$, and $m'' : {\cal I}_{\cal L} \rightarrow {\cal 
I}_{\cal B}$.

We summarize our requirements:

\begin{enumerate}

\item $L$ is a one-to-one map

\item at the end of filling up the first $C$ buckets, the number of tokens in 
each bucket can differ at most by 1 from the number of tokens in every 
other bucket

\item let $p(q)$ be the number of tokens with $q=l(t)\ mod B$, at the end of 
filling up the first $C$ buckets each $p(q)$ can differ at most by 1 from 
every other $p(q')$

\item to fill up the $B-C$ empty buckets, tokens can be moved only once 
from the $C$ buckets to the $B-C$ empty buckets, no redistribution 
between the $C$ buckets is allowed

\item at the end of redistributing the tokens in all the $B$ buckets, the 
number of tokens in each bucket can differ at most by 1 from the number 
of tokens in every other bucket and for all tokens in a bucket must 
hold $b=l(t)\ mod B$ where $b$ is the number of the bucket

\item after redistributing the tokens in the final $B'$ buckets, the number of 
tokens in each bucket can differ at most by 1 from the number of tokens 
in every other bucket and for all tokens in a bucket must hold 
$b'=l(t)\ mod B'$ where $b'$ is the number of the bucket.

\end{enumerate}

For implementational reasons, we add the following constraint on the solution:
of the two round-robin cycles of the first step, the
one of the tokens with $l(t)\ mod B$ not in ${\cal C}$ should run
in the direction of increasing values of $t$.

\section{Finding the solution}

First we need to determine the map $m$ in the general case. It is 
obvious that $m(t)=l(t)\ mod B$ if $C=B$. In the general case, if 
$t+f\ mod B\ \in\ {\cal I}_{\cal C}$ then we set $m(t)=l(t)\ mod B$. 
This already satisfies part of requirement 5. The simplest choice of $l(t)$ 
would be $l(t)=t+f$. But we should recall that we need two round-robin 
cycles running in opposite directions and that, due to the above mentioned
constraint, this round-robin cycle should run in the opposite 
direction of increasing values of $t$. 
Pratically this means that we would like to have something like 
$l(0)=f+C-1$, $l(1)= f+C-1-1$, ..., 
$l(C-1)=f$; that is something like $l(t)=f+C-1-t$. But this obviously 
cannot be right since it contains $-t$ and then $l$ would decrease 
steadily instead of increasing in jumps. 

So we should use instead a more complicated function which should 
satisfy: $l(0)=f+C-1$, $l(1)= f+C-1-1$, ..., $l(C-1)=f$, and then 
$l(B)=f+B+C-1$, $l(B+1)=f+B+C-1-1=f+(B+1)+C-1-2$, ..., 
$l(B+C-1)=f+B=f+(B+C-1)+C-1-2(C-1)$ which is given by 
$l(t)=f+t+C-1-2*(t\ mod B)$.

Now with this definition of $l(t)$, the map $m(t)=l(t)\ mod B$
puts a token with label $l$ in the bucket 
number $m$ going from $C-1$ to $0$ when $t$ goes from $0$ to $C-1$. 
This works for all tokens such that $f+t\ mod B\ \in\ {\cal I}_{\cal C}$.

For the tokens that should go in the $B-C$ buckets we are not filling in 
the first round, we adopt a simple round-robin scheduling. We set 
$l(t)=f+t$ and use a counter $r$ which increases independently every 
time a token is put in one of the $C$ buckets. The counter $r$ is 
an integer from $0$ to $C-1$ $mod C$.
In practice we put the token $C$ in the bucket $f$ and increase $r$ 
by $1$, the token $C+1$ in the bucket $f+1$ ... the token $B-1$ 
in the bucket $f+B-1-C$ (all $mod B$) always increasing $r$ by 
$1$\ $mod C$. After that we go back to the first cycle since for the
next token it holds  $f+t\ mod B\ \in \ {\cal I}_{\cal C}$.
When we get to the token $t=B+C$, the next one in the second cycle, we
should put it in the next bucket pointed to by $r$, and continue like that.

After this, we claim that the two next shuffling of tokens are simply 
given by the maps: $m''(t)=l(t)\ mod B$ and $m'(t) = l(t) \ mod B'$.

Let's summarize the solution we found. To specify the solution we need
the definition of four maps: $m(t)$, $m''(t)$, $m'(t)$ and $l(t)$. 
We have

\[ 
\begin{array}{ll} 
 m(t) &= \left\{ \begin{array}{ll} 
    l(t)\ mod B & \mbox{if $f+t\ mod B\ \in\ {\cal I}_{\cal C}$} \\
    r           & \mbox{otherwise}
                  \end{array}
          \right. \\
 & \\
 m''(t) &= l(t)\ mod B \\ 
 & \\
 m'(t) &= l(t) \ mod B' \\
 & \\
 l(t) &= \left\{ \begin{array}{ll} 
  f+t+C-1-2*(t\ mod B) & \mbox{if $f+t\ mod B\ \in\ {\cal I}_{\cal C}$} \\
  f+t                  & \mbox{otherwise}
                  \end{array}
          \right. \\
\end{array}
\]

\section*{Verifying the solution}

To check if this solution satisfies our requirements (we are not going
to make here a formal mathematical proof, even if it is not too difficult
but it is lenghty and not too much illuminating) we will show how each
of the steps of distributing the tokens in the buckets works.

First of all, requirement 1., $L$ is a one-to-one map, is automatically
satisfied by the definition of the map.

In the first step we put tokens only in the $C$ buckets of the ${\cal C}$
set. The condition $f+t\ mod B\ \in\ {\cal I}_{\cal C}$ and the 
round-robin distribution of the other tokens, guarantee that all 
tokens are put only in the buckets of the ${\cal C}$ set. 
Moreover the buckets
with $f+t\ mod B\ \in\ {\cal I}_{\cal C}$ are put in the correct 
bucket also for the second distribution, that is they will not be moved
in the second distribution.

Let's consider requirement 2. When $f+t\ mod B\ \in\ {\cal I}_{\cal C}$
we put the tokens in the buckets $f+C-1$ $..$ $f$. When we finished 
the first $C$ tokens, we start putting the next $B-C$ tokens in $f$,
$f+1$ etc. When we finish the tokens with 
$f+t\ mod B\ \not\in\ {\cal I}_{\cal C}$ 
we have in general at most a difference of $1$ in the
number of tokens in the buckets of the ${\cal C}$ set. 
Then we start again with
the first cycle and we put exactly one token per bucket in the ${\cal C}$
set. And we continue like this. The point to discuss is what happens 
at the end.

If the last token has $f+t\ mod B\ \not\in\ {\cal I}_{\cal C}$ then
for what we just said the difference in number of tokens in the
buckets of the ${\cal C}$ set is at most $1$. If instead 
$f+t\ mod B\ \in\ {\cal I}_{\cal C}$ 
then the situation is more complex. If we put
the last token in the $f$ bucket, then again the difference in number 
of tokens in the buckets of the ${\cal C}$ set is at most $1$ as before. 

Otherwise denote by $z$ the bucket in which we put the last token of 
the second (round-robin) cycle and $y$ the bucket in which we put 
the very last token. If $z+1=y$ (remember the direction of filling
of the two cycles) then all buckets in the ${\cal C}$ set 
have the same number of tokens.
If $z+1<y$ then the buckets in the ${\cal C}$ set 
with number in between $z$ and $y$
have one token less than the others. If instead $z+1>y$ then the buckets 
in the ${\cal C}$ set with number in between $z$ and $y$ have one token more 
than the others.

Notice that in the case in which the last token is in the first cycle
and is not in the $f$ bucket, the list of integer numbers
$l(t)$ has a gap. Indeed $l(t)$ starts at $f$, but if we write 
the last value as $N*B+M$ then the values $N*B$, $N*B+1$, $...$, 
$N*B+M-1$ are missing from the ${\cal I}_{\cal L}$ set.

In conclusion the difference in number of tokens in the ${\cal C}$ set
is always at most 1. This shows that requirement 2.\ is satisfied.

Requirement 3.\ is again automatically satisfied by our definition of 
$L$.

From the definition of $l$ and $m''$ it follows that only the tokens
distributed in the second round-robin cycle are moved in the second
step, they are moved once and directly to their final bucket in the
${\cal B}$ set. This shows that requirement 4.\ is satisfied. 

Since $L$ is a one-to-one map and from the definition of the $m''$ map
it follows that also requirement 5.\ is satisfied.

Finally again due to the definition of $L$ and of the $m'$ map, also
requirement 6.\ is satisfied.

\section*{Conclusive remarks}
The solution we described for the problem at hand is {\sl one} solution,
most probably it is not the only one. For example, without the 
constraint on the direction of the two round-robin cycles, we could
choose the opposite direction for them and get a different solution.

Moreover, we do not claim that this is the simplest or more efficient
solution, but it is one which practically was easy to implement using
a couple of pointers.

\section*{Acknowledgments}
We thank H.\ Bechmann-Pasquinucci for inspiring remarks.

\end{document}